%slight corrections 29th October 1996
\input harvmac.tex
\input epsf.tex
\parindent=0pt
\parskip=5pt

\hyphenation{satisfying Kronheimer}

\def\IR{{\hbox{{\rm I}\kern-.2em\hbox{\rm R}}}}
\def\IB{{\hbox{{\rm I}\kern-.2em\hbox{\rm B}}}}
\def\IN{{\hbox{{\rm I}\kern-.2em\hbox{\rm N}}}}
\def\IC{\,\,{\hbox{{\rm I}\kern-.59em\hbox{\bf C}}}}
\def\IZ{{\hbox{{\rm Z}\kern-.4em\hbox{\rm Z}}}}
\def\IP{{\hbox{{\rm I}\kern-.2em\hbox{\rm P}}}}
\def\IH{{\hbox{{\rm I}\kern-.4em\hbox{\rm H}}}}
\def\ID{{\hbox{{\rm I}\kern-.2em\hbox{\rm D}}}}
\def\mmu{{\mu}\hskip-.55em{\mu}}

\def\N{{\cal N}}
\def\T{{\cal T}}
\def\O{{\cal O}}
\def\I{{\cal I}}

\def\Tr{{\rm Tr}}

\noblackbox

\Title{\vbox{\baselineskip12pt
\hbox{NSF-ITP-96-140}
\hbox{McGill/96-32}
\hbox{hep-th/9610140}}}
{Aspects of Type~IIB Theory on ALE Spaces}

\centerline{ \bf Clifford V. Johnson$^a$ and Robert C.~Myers$^b$ }
\bigskip\centerline{$^a${\it Institute for Theoretical Physics, UCSB,
CA~93106, USA }}
\bigskip\centerline{$^b${\it Physics Department, McGill
University, Montr\'eal, PQ, H3A 2T8 Canada}}
\footnote{}{\sl email: $^a${\tt cvj@itp.ucsb.edu},  
$^b${\tt rcm@hep.physics.mcgill.ca}}
\vskip1.5cm
\centerline{\bf Abstract}
\vskip0.7cm
\vbox{\narrower\baselineskip=12pt\noindent
D--brane technology and strong/weak coupling duality supplement
traditional orbifold techniques by making certain background
geometries more accessible.  In this spirit, we consider some of the
geometric properties of the type~IIB theory on $\IR^6{\times}{\cal M}$
where ${\cal M}$ is an `Asymptotically Locally Euclidean (ALE)'
gravitational instanton. Given the self--duality of the theory, we can
extract the geometry (both singular and resolved) seen by the weakly
coupled IIB string by studying the physics of a D1--brane probe. The
construction is both amusing and instructive, as the physics of the
probe completely captures the mathematics of the construction of ALE
instantons via `HyperK\"ahler Quotients', as presented by
Kronheimer. This relation has been noted by Douglas and Moore for the
$A$--series. We extend the explicit construction to the case of the
$D$-- and $E$--series --- uncovering a quite beautiful structure ---
and highlight how all of the elements of the mathematical construction
find their counterparts in the physics of the type~IIB D--string.  We
discuss the explicit ALE metrics which may be obtained using these
techniques, and comment on the role duality plays in relating gauged
linear sigma models to conformal field theories.}
\vskip0.5cm

%\draft
\Date{October 1996}

\baselineskip13pt

\lref\dbranes{J.~Dai, R.~G.~Leigh and J.~Polchinski, 
{\sl `New Connections Between String Theories'}, Mod.~Phys.~Lett.
{\bf A4} (1989) 2073\semi P.~Ho\u{r}ava, {\sl `Background Duality of
Open String Models'}, Phys. Lett. {\bf B231} (1989) 251\semi
R.~G.~Leigh, {`Dirac--Born--Infeld Action from Dirichlet Sigma
Model'}, Mod.~Phys.~Lett. {\bf A4} (1989) 2767\semi J.~Polchinski,
{\sl`Combinatorics Of Boundaries in String Theory'}, Phys.~Rev.~D50
(1994) 6041, hep-th/9407031.}

\lref\mumford{D. Mumford and J. Fogarty, {\sl `Geometric Invariant Theory'},
Springer, 1982.}
\lref\aspinwallone{See for example, P. Aspinwall, {\sl `Resolution of
Orbifold Singularities in String Theory'}, in `Essays On Mirror Manifolds 
2', hep-th/9403123, and references therein.}
\lref\aspinwalltwo{Paul Aspinwall, {\sl `Enhanced Gauge Symmetries and $K3$ 
Surfaces'}, Phys. Lett. {\bf B357} (1995) 329, hep-th/9507012.}
\lref\edcomm{E. Witten, {\sl`Some Comments On String Dynamics'}, in the 
Proceedings of {\sl Strings 95}, USC, 1995, hep-th/9507121.}
\lref\joetensor{J. Polchinski, {\sl`Tensors From $K3$ Orientifolds'}, 
hep-th/9606165.}
\lref\kronheimer{P. B. Kronheimer, {\sl `The Construction of ALE Spaces as 
Hyper--K\"ahler Quotients'}, J.~Diff. Geom. {\bf 29} (1989) 665.}
\lref\hitchin{N. J.  Hitchin,  {\sl `Polygons and Gravitons'}, Math. Proc. 
Camb. Phil. Soc. {\bf 85} (1979) 465.}
\lref\moore{M. R. Douglas and G. Moore,  {\sl`D--Branes, Quivers and ALE 
Instantons'}, hep-th/9603167.}
\lref\rocek{N. J. Hitchin, A. Karlhede, U. Lindstr\"om and M. Ro\u cek, 
{\sl `Hyper--K\"ahler Metrics and Supersymmetry'}, Comm. Math. Phys. {\bf 108}
(1987) 535.}
\lref\klein{F. Klein, {\sl`Vorlesungen \"Uber das Ikosaeder und die 
Aufl\"osung der Gleichungen vom f\"unften Grade'}, Teubner, Leipzig 1884;
F. Klein, {\sl `Lectures on the Icosahedron and  the Solution of an Equation 
of Fifth Degree'}, Dover, New York, 1913.}
\lref\elliot{J. P. Elliot and P. G. Dawber, {\sl `Symmetry in Physics'}, 
McMillan, 1986.}
\lref\ericjoe{E. G. Gimion and J. Polchinski, {\sl`Consistency
 Conditions of Orientifolds and D--Manifolds'}, Phys. Rev. {\bf D54} (1996) 
1667, hep-th/9601038.}
\lref\ericmeI{E. G. Gimion and C. V. Johnson, {\sl`$K3$ Orientifolds'}, Nucl. 
Phys. {\bf B478} (1996), hep-th/9604129.}
\lref\ericmeII{E. G. Gimion and C. V. Johnson, {\sl`Multiple Realisations of 
${\cal N}{=}1$ Vacua in Six Dimensions'}, Nucl. Phys. {\bf B479} (1996), 285,
hep-th/9606176}
\lref\McKay{J. McKay, {\sl `Graphs, Singularties
 and Finite Groups'}, Proc. Symp. Pure. Math. {\bf 37} (1980) 183,
Providence, RI; Amer. Math. Soc.}
\lref\orbifold{L. Dixon, J. Harvey, C. Vafa and E. Witten, {\sl `Strings on 
Orbifolds'}, Nucl. Phys. {\bf B261} (1985) 678;
{\it ibid}, Nucl. Phys. {\bf B274} (1986) 285.}
\lref\algebra{P. Slodowy, {\sl`Simple Singularities and Simple Algebraic 
Groups'}, Lecture Notes in Math., Vol.  {\bf 815}, Springer, Berlin, 1980.}
\lref\gibhawk{G. W. Gibbons and S. W. Hawking, {\sl `Gravitational
Multi--Instantons'}, Phys. Lett. {\bf B78} (1978) 430.}
\lref\eghan{T. Eguchi and A. J. Hanson, {\sl`Asymptotically Flat 
Self--Dual Solutions to Euclidean Gravity'}, Phys. Lett. {\bf B74} (1978) 249.}
\lref\adhm{E. Witten, {\sl `Sigma Models and the ADHM Construction of 
Instantons'}, J.~Geom.  Phys. {\bf 15} (1995) 215, hep-th/9410052.}
\lref\small{E. Witten, {\sl `Small Instantons in String Theory'},  Nucl. 
Phys. {\bf B460} (1996) 541, hep-th/9511030.}
\lref\douglas{M. R.  Douglas, {\sl `Gauge Fields and D--Branes'},  
hep-th/9604198.}
\lref\phases{E. Witten, {\sl `Phases of $N{=}2$ Theories in Two Dimensions'}, 
Nucl. Phys. {\bf B403} (1993) 159,  hep-th/9301042.}
\lref\edbound{E. Witten, {\sl `Bound States of Stirngs and $p$--Branes'}, 
Nucl. Phys. {\bf B460} (1996) 335, hep-th/9510135.}
\lref\ADHM{M. F. Atiyah, V. Drinfeld, N. J. Hitchin and Y. I. Manin, {\sl
`Construction of Instantons'} Phys. Lett. {\bf A65} (1978) 185.}
\lref\kronhiemerii{P. B. Kronheimer and H. Nakajima, {\sl `Yang--Mills 
Instantons on ALE Gravitational Instantons'}, Math. Ann. {\bf 288} (1990) 263.}
\lref\italiansi{M. Bianchi, F. Fucito, G. Rossi, and M. Martinelli, 
{\sl `Explicit Construction of Yang--Mills Instantons on ALE Spaces'}, 
Nucl. Phys. {\bf B473} (1996) 367, hep-th/9601162.}
\lref\italiansii{D. Anselmi, M. Bill\'o, P. Fr\'e, L. Giraradello and A. 
Zaffaroni, {\sl `ALE Manifolds and Conformal Field Theories'},  Int. J. 
Mod. Phys. {\bf A9} (1994) 3007,  hep-th/9304135.}
\lref\nonrenorm{L. Alvarez--Gaume and D. Z. Freedman, {\sl `Geometrical 
structure and Ultraviolet Finiteness in the Supersymmetric Sigma Model'}, 
Comm. Math. Phys. {\bf 80} (1981) 443.} 
\lref\myoldpaper{C. V. Johnson, {\sl `Exact Models of Extremal Dyonic 4D 
Black Hole Solutions of Heterotic String Theory'}, Phys. Rev. {\bf D50} (1994)
4032, hep-th/9403192.}
\lref\gojoe{J. Polchinski, {\sl `Dirichlet Branes and Ramond--Ramond Charges
 in String Theory'}, Phys. Rev. Lett. {\bf 75} (1995) hep-th/9510017.}
\lref\dnotes{J. Polchinski, S. Chaudhuri and C. V. Johnson, {\sl `Notes on 
D--Branes'}, hep-th/9602052.}
\lref\hulltown{C. M. Hull and P. Townsend, {\sl `Unity of Superstring 
Dualities'}, Nucl. Phys. {\bf B438} (1995) 109, hep-th/9410167.}
\lref\hull{C. M.  Hull, {\sl `String--String Duality in Ten Dimensions'}, 
 Phys. Lett. {\bf B357} (1995) 545,  hep-th/9506194.}
\lref\ken{K. Intriligator and  N. Seiberg,  {\sl `Mirror Symmetry in Three 
Dimensional Gauge Theories'}, hep-th/9607207.}

%%%%%%%%%%%%%%%%%%%%%%%%%%%%%%%%%%%%%%%%%%%%%%%%%%%%%%%%%%%%%%%%%%%%%%%%%%

\newsec{Introduction}

\subsec{\sl Motivation}
The self--duality of the ten dimensional type IIB theory under the
strong/weak coupling duality map\hulltown\ makes it a particularly
interesting theory to study. The use of D--brane
technology\refs{\dbranes,\gojoe,\dnotes}\ as an aid in gaining insight
into the physics of various regimes of the theory makes its study
somewhat more tractable.

Of particular interest for us will be the D1--brane. This
soliton--like BPS saturated object has the distinction that it gets
exchanged with the `fundamental' type~IIB string under the strong/weak
coupling duality transformation, becoming the `fundamental' string of
the dual theory. The self--duality of the theory is seen here by
noting that the zero--mode spectra of both strings are identical\hull.

One might expect then, that aspects of the type~IIB theory which might
seem obscure or difficult to handle by studying the spectrum of the
`fundamental' string might be better addressed in studying its dual
partner, the D1--brane. The converse is also to be expected.

While this is true for any dual pair of theories, the novelty here is
that for the {\it self--dual} IIB theory, these are dual descriptions
of physics which is still essentially perturbative from the point of
view of {\it either} string theory. The dual descriptions are thus
rather more like a conventional change of variables than
usual\foot{This additional aspect of IIB strong/weak coupling duality
does not make it any less profound than the other dualities, from a
number of points of view.}.

It is with this in mind that we examine the physics of the type~IIB
theory in the neighbourhood of an ALE gravitational instanton, using a
D1--brane probe\douglas. This is a particularly interesting set of
backgrounds to study, for many reasons. Chief among those, in this
context, are:

{\sl (i)} their blow--down limits are simple to describe as an
orbifold\italiansii;\hfill\break {\sl (ii)} they are completely
classified\hitchin\ (falling into an $A$--$D$--$E$
series);\hfill\break {\sl (iii)} as string backgrounds they break only
half of the supersymmetry; and
\hfill\break
 {\sl (iv)} they are non--compact spaces, thereby allowing us to study
them in the context of self--dual type~IIB theory in ten dimensions.

\subsec{\sl Orbifolds, ALE spaces and Geometry}
Much of the physics of strings on ALE spaces was well understood using
orbifold technology\italiansii. One studies the string theory on the
singular space $ \IR^4/\Gamma$ where $\Gamma$ is some discrete
subgroup of $SU(2)$ (they fall into an $A$--$D$--$E$
classification\klein) acting on the space $\IR^4$. The result that
string theory is naturally well behaved on such a singular space is
related to the appearance (required by modular invariance) of massless
states from `twisted sectors' of the orbifold\orbifold, which
correspond to precisely the moduli needed to deform the theory to the
neighbouring problem with a smooth target space. So even without
knowledge of the detailed form of the metric for the ALE spaces it is
enough to know that the stringy resolution is complete by
comparing\aspinwallone\ to (say) a purely algebraic construction of
the moduli space of the objects which form the target space.

So in principle, the string theory `knows' everything about the metric
on the ALE spaces. As these spaces are quite simple and well--studied
(see later for a review), there is apparently not much in the way of
new physics to discover, at least away from the regimes of the theory
where the  new (and, currently, poorly understood) phenomena first
characterised in ref.\refs{\edcomm}\ arise. The description via
orbifolds is very adequate in capturing the physics.

However, there is a matter of both principle and practice here.
First, it is difficult (at best) to extract the detailed form of the
metric that the string sees, using the orbifold technology. More
generally, we can usually extract the metric of a target space defined
by a conformal field theory only when we have a Lagrangian definition
of it, such as a (gauged or ungauged) Wess--Zumino--Witten
model. Unfortunately, such path integral definitions are only known
for a very small subset of the conformal field theories of interest.
Furthermore, even in principle it is not clear how the string feels
its way around the smooth space on which the complete orbifold
suggests that it is propagating.

This is one of the issues we would like to highlight in this
paper. The techniques described herein are a means of extracting
directly from the string theory itself (using strong/weak coupling
duality and D--branes) the details (not just algebraic, but
differential) of the resolved spaces which the orbifolded fundamental
string theory sees.

Most often, we study string theory on a target space of choice by
putting the metric into the formalism ourselves.  In practice, we
usually proceed to find the metric on the target space by solving the
background field equations by hand (order by order in
$\alpha^\prime$), aided by the string theory only in the cases where
some symmetry of the theory provides us with a solution generating
technique, or some other such means of making the problem more
manageable. In general though, in order to find the solutions of
interest we have to employ methods which are often supplementary to
string theory itself.

In the case of the ALE instantons, which fall into an $A$--$D$--$E$
classification\hitchin, only the metric for the $A_k$ series is known in a
closed expression:
\def\y{{\bf y}}
\def\A{{\bf A}}
\eqn\Ametric{\eqalign{&ds^2=V^{-1}(dt-\A\cdot d\y)^2+V d\y\cdot d\y\cr
&{\rm where}\quad V=\sum_{i=1}^{k}{1\over |\y-\y_i|}\cr
&{\rm and}\quad \nabla V=\nabla\times\A.}}

These are the Gibbons--Hawking multi--centre metrics\gibhawk. The case
$k{=}1$ is the Eguchi--Hanson metric\eghan.  These spaces are
asymptotically flat, but Euclidean only locally: There is a global
identification which makes the surface at infinity $S^3/\IZ_{k+1}$
instead of $S^3$.

As solutions of the vacuum Einstein equations, these metrics are also
solutions of the leading--order background field equations in type~IIB
string theory\foot{In fact, these backgrounds are solutions to all
orders in the $\alpha'$ expansion because the corresponding
world--sheet theory has $\N=4$ supersymmetry\nonrenorm.}.  However,
their derivation as solutions of Einstein's equations were not
particularly stringy in origin.

For a time after these metrics were found, the instantons
corresponding to the $D$ and $E$ series were not known, although their
existence was strongly motivated in ref.\hitchin.  For the $A_k$
series, the ALE nature arises from the discrete identifications
$\IZ_{k+1}$ at infinity. This is the cyclic subgroup of $SU(2)$. For
the $D_k$ and $E_{6,7,8}$ series, the cyclic group is replaced by 
the binary dihedral ($\ID_{k-2}$), tetrahedral ($\T$), octahedral
($\O$) and icosahedral ($\I$) groups.  (We will remind the reader of
the relation\klein\ between the discrete subgroups of $SU(2)$ and the
simply laced Lie algebras later in this paper.)

\subsec{\sl HyperK\"ahler Quotients}
In 1987, an explicit construction of the spaces (and a proof of the
conjectured $A$--$D$--$E$ classification) was presented by
Kronheimer\kronheimer. The main tool used was the `HyperK\"ahler
Quotient' technique\rocek. In short (more details later) it was shown
that one can recover the ALE spaces by starting with a parent space
$M$ (a flat hyperK\"ahler manifold of high dimension) acted on by some
auxiliary group $F$. By virtue of the hyperK\"ahler structure of~$M$,
the group action naturally induces a Lie--algebra--valued
triplet of functions 
${\mu}\hskip-.55em{\mu}$ called the `moment map', which plays a
central role in the construction. It in turn defines naturally a set
of constraints and a coset procedure which recover a space of real
dimension four, possessing the properties of the sought--after ALE
spaces, for the appropriate choice of $F$ and $M$.

The
resulting spaces are again solutions of the IIB string theory by
virtue of their being solutions of Einstein's equations for empty
space.

Once again, the techniques used to find the solutions lie outside
(apparently, as we shall see) the realm of string theory, using as
input a number of auxiliary objects such as the group $F$ and the
parent hyperK\"ahler manifold $M$. 

\subsec{\sl The Dual Picture}
Amusingly, while for the traditional orbifold description of the IIB
theory on an ALE space, the above paragraph is true, in the dual
picture --- described by the D1--brane --- the physics of how the
string sees the metric can be made explicit, and it maps onto
precisely Kronheimer's construction! This was shown for the
$A$--series in ref.\moore. This paper extends the result to the full
$A$--$D$--$E$ family.

In particular, there is a physical role for the group $F$, the parent
space $M$ and the moment map ${\mu}\hskip-.55em{\mu}$, which were all
essential ingredients of the mathematical construction.

The world--volume theory of the fundamental IIB string on the ALE
space ---an $\N{=}4$ conformal field theory--- is exchanged under
duality for the world--volume theory of the D1--brane probe, which is
an $\N{=}4$ gauged linear sigma model. The gauge group is $F$, which
acts on a set of hypermultiplet scalars parameterising a manifold
$M$. The allowed values of the hypermultiplets parameterise the
position of the D1--brane probe in space. Solving the D--flatness
conditions for the allowed values of the hypermultiplets is equivalent
to solving the constraints imposed by the moment
map~${\mu}\hskip-.55em{\mu}$.  The metric on that moduli space of
vacua is the spacetime metric seen by the D1--brane. As the type~IIB
theory is self--dual, we have therefore learned how the orbifolded
fundamental type~IIB string actually recovers metric data about the
smooth space it propagates in.

So it seems that we have come full circle: In order to find the metric
that the string sees on the ALE spaces we move progressively further
away from stringy techniques to solve the problem.  In doing so, we
`meet ourselves coming the other way', working deep inside the string
theory, probing with D--brane variables!

\subsec{\sl Linear Sigma Models and D--Branes}
The fact that gauged linear sigma models capture the essence of
certain quotient constructions used in mathematics and mathematical
physics has been noticed and exploited successfully in the string
theory context\phases. The essential physical picture of the
hyperK\"ahler quotient construction and its relevance to strings
propagating near ALE singularities was discussed in detail in
ref.\edcomm\ for the $A_1$ singularity.  These discussions all took
place before the recognition\gojoe\ of the role D--branes to duality
and related issues\foot{The reader may also wish to consult
ref.\italiansii\ for a very useful review of Kronheimer's
hyperK\"ahler quotient construction of ALE spaces. Again, the quotient
construction is treated there as supplementary to the string
theory. The paper's main concern is a construction of the relevant
orbifold conformal field theories and the study of their marginal
deformations.}.

That D1--branes are a natural means of producing two dimensional
gauged linear sigma models was noticed in ref.\edbound.
Douglas\douglas\ made use of this technique
in the type~I context to recover ({\it
via} duality) the linear sigma models associated to heterotic strings
in Yang--Mills instanton backgrounds\adhm, pioneering the explicit use
of D1--branes as probes of short--distance 
string physics.  There
D--brane and duality technology captured the ADHM hyperK\"ahler
quotient construction of Yang--Mills instantons\ADHM. (The Yang--Mills
instantons live at the core of heterotic fivebranes, which are dual to
type~I's D5--branes in the small instanton limit.)

The D1-- and D5--brane constructions in ref.\douglas\ were further
extended to include Yang--Mills instantons on the $A$--series ALE
spaces in ref.\moore\ where it was observed that the hyperK\"ahler
quotient construction of Kronheimer and Nakajima\kronhiemerii\ was
this time cast into a physical setting. 

In a different but related context, the explicit details of the
D1--branes on the A--series ALE spaces were worked out in
ref.\joetensor. There, the exercise (carried out for the $k{=}1$ case
in for example ref.\italiansi) of choosing good coordinates on the
space of hypermultiplets and deriving the form \Ametric\ as the metric
on moduli space is carried out in a D--brane context for the full
$A_k$ series.

\subsec{\sl Outline}
In this paper we shall extend the explicit construction of D1--brane
physics on the ALE spaces to the complete ($A$--$D$--$E$) family of
ALE spaces. Along the way we find that the D--brane physics is a
remarkably clear and simple guide to the elements of the mathematical
construction. At the same time, many beautiful mathematical results
are united in this simple physical setting. The results make the
potentially unpleasant structures present in the $D$ and $E$ series
(due to the non--linearities introduced by the non--Abelian nature of
$\Gamma$) more manageable in many respects, while allowing the elegance
of the $A$--$D$--$E$ classification to shine throughout.

In section~2 we start the study of strings on ALE spaces by first
considering the singular `blow--down' limit.  We first introduce the
discrete subgroups $\Gamma$ of $SU(2)$ and their natural action on
$\IR^4$. Introducing the D1--brane probe(s) we impose a projection on
the Chan--Paton factors to ensure that they respect the $\Gamma$
symmetry, and solve for the spectrum of the the world--volume
theory. Using the  result of McKay\McKay\ concerning the
representations of $\Gamma$ we find that the D--branes are organised
according to the structure of the extended Dynkin diagrams associated
to the $A$--$D$--$E$ root systems. The resulting spectrum of the
world--volume theory ($D{=}2$, $\N{=}4$) is described. We pause to
note the relation to  elements of the construction of
Kronheimer.

A discussion of the various branches of the classical
moduli space of vacua of the world--volume theories maps out the
geometry that the probes see, simultaneously introducing the essence
of the hyperK\"ahler quotient construction of the blown--down ALE
spaces.

In section~3 we complete the construction, by turning on the closed
string fields which deform to resolved ALE spaces. After discussing
how these couple in the world--volume theory (following ref.\moore)
we revisit the Higgs branch of the moduli space of vacua and complete
the discussion of IIB strings on resolved ALE spaces, thus also
completing the hyperK\"ahler quotient discussion of Kronheimer.

A discussion of the explicit metric on moduli space ({\it i.e.},
explicit metrics on the ALE spaces) takes place in section~4, and with
a few remarks and speculations in section~5 we end this presentation.

\newsec{D--Strings on ALE Spaces: The Blow--Down}
\subsec{\sl Preliminaries}
Starting at an arbitrary point in the moduli space of an ALE
instanton, we can adjust certain moduli in order to shrink the
`core' of the instanton, locating the associated curvature in a
progressively smaller region of four dimensional Euclidean space.  The
limit of this procedure is the `blow--down' or `orbifold limit' of the
space, when all of the curvature is localised at a single singular
point. The blown--down ALE instanton is essentially just the space
$\IR^4/\Gamma$ where $\Gamma$ is a discrete group. (For simplicity,
we'll place the singular point at the origin of our coordinates, in
our definitions which follow.)

Let us denote the Cartesian coordinates on the $\IR^4$ as $x^6,x^7,x^8$
and $x^9$.  Let us also define a  set of complex coordinates, 
$z^1{=}x^6{+}ix^7$ and $z^2{=}x^8{+}ix^9$. Thus we will
sometimes describe the space as the complex space $\IC^2$,
coordinatised by the $z^i$. Many of the results of our analysis will
inherit this structure.

Yet another natural description of $\IR^4\equiv\IC^2$ is as
a trivial quaternionic space, where we can write the coordinate
\eqn\quater{q=\pmatrix{z^1&-{\bar z}^2\cr z^2 &\phantom{-}{\bar z}^1}.}
There is an $SU(2)_L{\times}SU(2)_R\subset SO(4)$ group action on the space
given by 
\eqn\suaction{q\to g_L\cdot q\cdot  
g_R,\quad{\rm for}\,\,\,g_{L,R}\in SU(2)_{L,R}.}  Choosing either of
the $SU(2)$'s (we choose the left), the orbits of its free action on
the $z^i$ parameterise families of $S^3$'s, the natural invariant
submanifolds of the action of $SO(4)$ on~$\IR^4$. The $SU(2)_R$ will
remain as the standard global symmetry associated with hyperK\"ahler
manifolds (eventually manifesting itself as an R--symmetry in the
supersymmetric field theories which we study later). 

The asymptotic group of discrete identifications, $\Gamma$, of the ALE
spaces are subgroups of $SU(2)$. These groups have been classified by
Klein\klein. They are governed by the same Diophantine equations which
organise the point group symmetries of regular solids in three
dimensions\foot{See for example ref.\elliot\ for a nice description of
the point groups, and ref.\algebra\ for a review and discussion of the
Kleinian singularities.}, and consequently fall into an $A$--$D$--$E$
classification.

There are\algebra:

\def\A{{\cal A}}
\def\B{{\cal B}}

\def\N{{\cal N}}

(i) The $A_k$ series ($k{\geq}1$). This is the set of cyclic groups of
order $k{+}1$, denoted $\IZ_{k+1}$.  Their action on the $z^i$ is
generated by
\eqn\Agener{g=\pmatrix{e^{2i\pi\over k+1}&0\cr0 & e^{-{2i\pi\over k+1}}}.}

(ii) The $D_k$ series ($k{\geq}4$). This is the binary extension of
the dihedral group, of order $4(k{-}2)$, denoted $\ID_{k-2}$.  Their
action on the $z^i$ is generated by
\eqn\Dgener{\A=\pmatrix{e^{i\pi\over k-2}&0\cr0 
&e^{-{i\pi\over k-2}}}\quad{\rm and}\quad
\B=\pmatrix{0&i\cr i&0}}
In this representation the central element is ${\cal
Z}{=}{-}1({=}\A^2{=}\B^2{=}(\A\B)^2)$.  Note that the generators $\A$
form a cyclic subgroup $\IZ_{2k-4}$.

(iii) The remaining groups fall into the $E_{6,7,8}$ series, and are
the binary tetrahedral ($\T$), octahedral ($\O$) and icosahedral
($\I$) groups of order 24, 48 and 120, respectively.

The group $\T$ is generated by taking the elements of $\ID_2$ and
combining them with
\eqn\Egener{{1\over\sqrt2}\pmatrix{\varepsilon^7&\varepsilon^7\cr
\varepsilon^5 
&\varepsilon}} where $\varepsilon$ is an 8th root of unity.

The group $\O$ is generated by taking the elements of $\T$ and
combining them with
\eqn\Egener{\pmatrix{\varepsilon&0\cr0
&\varepsilon^7}.}
Finally $\I$ is generated by
\eqn\Dgener{-\pmatrix{\eta^3&0\cr0
&\eta^2}\quad{\rm and}\quad{1\over{\eta^2{-}\eta^3}}\pmatrix{\eta{+}
\eta^4&1\cr
1&{-}\eta{-}\eta^4},}  where $\eta$ is a 5th root of unity.

More properties of all of these discrete groups will appear as we
proceed.  Let us postpone their introduction until such time as we
need them, and now turn to the string theory.

\subsec{\sl The Open String Sector}
We start by  considering string propagation on $\IR^6{\times}\IR^4/\Gamma$.
 The orbifold space breaks half of the $D{=}10$,
$\N{=}2$ supersymmetry, leaving us with (thinking of this as a
compactification) $\N{=}2$ in the six dimensions of the $\IR^6$, with
coordinates $x^0,\ldots,x^5$.

We introduce a family of parallel D1--brane probes, all lying in (say)
the $x^0, x^1$ directions, and positioned in the $x^6,\ldots,x^9$
space in a $\Gamma$--invariant way. These break half of the
supersymmetry again, leaving us with $\N{=}1$ in six dimensions. We
shall focus on the world--sheet theories of the D1--branes
(D--strings) which are thus $D{=}2$ field theories with $\N{=}4$
supersymmetry, via dimensional reduction.

Our spacetime is non--compact everywhere, and so it is consistent to
place an arbitrary number of these branes in the
problem\dnotes\foot{See for example refs.\refs{\ericjoe,\ericmeI}\ for
studies of D--branes near ALE singularities in the context of compact
internal spacetimes, the manifold $K3$. There, the number of branes is
fixed.}.  However, we should ensure that there are enough D--branes in
the problem to ensure a faithful representation of the discrete
group~$\Gamma$ in the open string sector\ericmeI.  We therefore
introduce $|\Gamma|{=}k{+}1,4(k{-}2),24,48$ or $120$ D--strings
depending upon whether we are studying the $A_{k}$, $D_{k}$ or
$E_{6,7,8}$ series.

We will ask that the group $\Gamma$ be represented on the D--brane
(Chan--Paton) indices by the action of $|\Gamma|{\times}|\Gamma|$
matrices $\gamma^{\phantom{-1}}_\Gamma\!\!.$ This is the `regular'
representation.  (In what follows, appearances of
$\gamma^{\phantom{-1}}_\Gamma\!\!$ will be taken to imply the action
of all of the elements of the representation of $\Gamma$.)

Placing all of the D--branes at (say) the origin of the `internal'
$\IR^4\equiv\IC^2$, we have gauge group $U(|\Gamma|)$, arising in the
familiar way from massless open strings connecting the various coincident 
D--branes.
We then proceed by projecting this system to obtain invariance under the
group $\Gamma$. This will reduce the gauge group to some subgroup of
$U(|\Gamma|)$. Let us discover what this is.

There are three types of open string sector states 
to consider:

\bigskip

$\underline{\sl Vector-multiplets:}
\quad\quad\quad\lambda_V\,\psi_{-{1\over2}}^\mu|0\!> \quad\quad \mu=0,1$

The Chan--Paton matrix $\lambda_V$ starts out life as an arbitrary
$|\Gamma|{\times}|\Gamma|$ Hermitian matrix, a generator of the
$U(|\Gamma|)$ gauge group. Invariance under $\Gamma$ means that it
should satisfy additionally:
\eqn\vectss{\gamma^{\phantom{-1}}_\Gamma\!\!\!\lambda_V\gamma^{-1}_\Gamma
=\lambda_V.}  This constraint will give rise to vector fields (denoted
generically $A_\mu$) in the theory, transforming in the adjoint of
some subgroup of $U(|\Gamma|)$. We shall see what that subgroup is
shortly.

\bigskip

$\underline{{\sl Hyper-multiplets}\,\,{\sl  I:}}
\quad\quad\quad \lambda^{I}_H\,\psi_{-{1\over2}}^i|0\!> \quad\quad i=2,3,4,5$.

Here again the Chan--Paton matrices begin as arbitrary
$|\Gamma|{\times}|\Gamma|$ Hermitian matrix, giving hypermultiplets in
the adjoint of $U(|\Gamma|)$ (the rest of the $D{=}6$, $\N{=}1$
vectors from a $D{=}6$ point of view). They satisfy a similar equation
to
\vectss\ and result in hypermultiplets in the adjoint of the gauge 
group which results from \vectss. From the $\N{=}4$, $D{=}2$ point of
view, they are simply the scalar parts of the gauge multiplets.  These
scalars, which we shall denote $\phi^i_H$, parameterise motions of the
branes transverse to their world--volumes in the $x^2,x^3,x^4,x^5$
directions.

\vfill\eject

$\underline{{\sl Hyper-multiplets}\,\,{\sl II:}}
\quad\quad\quad\lambda^{II}_H\,\psi_{-{1\over2}}^m|0\!> \quad\quad m=6,7,8,9$.

In this sector, given the action of $\Gamma$ on $\IC^2$ as shown in
the previous subsection, it is prudent to relabel our string modes and
the resulting hypermultiplets to respect that structure. Our massless
modes are thus $\lambda^{1}_H\,\psi_{-{1\over2}}^1|0\!>$ and
$\lambda^{2}_H\,\psi_{-{1\over2}}^2|0\!>$, (and their adjoints) and
our constraint equation is more complicated than previously, as these
are the coordinates upon which the discrete group acts:
\eqn\hypers{\pmatrix{\gamma^{\phantom{-1}}_\Gamma\!\!\!\lambda^1_H
\,\gamma^{-1}_\Gamma\cr
\gamma^{\phantom{-1}}_\Gamma\!\!\!\lambda^2_H
\,\gamma^{-1}_\Gamma }=G_\Gamma\cdot\pmatrix{ \lambda^1_H\cr\lambda^2_H}}
where $G_\Gamma$ is a matrix acting in the $2{\times}2$ representation
of $\Gamma$, acting on the indices $j$ of the hypermultiplets
$\lambda^j_H$. We shall denote the resulting massless (complex) 
fields in the
world--volume theory as simply $\psi^1_H$ and $\psi^2_H$. (These are
not to be confused with the standard superconformal field theory modes
$\psi_{-{1\over2}}$. We shall not refer to those in what follows.) These
hypermultiplets parameterise the motion of the branes in the
$x^6,x^7,x^8,x^9$ directions.  In analogy with equation \quater,
these hypermultiplets are naturally
gathered together a quaternionic form as: 
\eqn\hyperquat{\Psi=\pmatrix{\psi^{1}_H&-\psi^{2\dagger}_H\cr
\psi^{2}_H&\phantom{-}\psi^{1\dagger}_H}.}

\subsec{\sl The $D{=}2$, $\N{=}4$ World--Volume Gauge Theory}
The quickest way to begin to see the solution to the equation \vectss\
is to note a little more of the structure of the discrete group
$\Gamma$ and its irreducible representations. In particular, recall
that elementary group theory tells us that we can always find a basis
in which the $|\Gamma|{\times}|\Gamma|$ `regular' representation of
$\Gamma$ is of block diagonal form\elliot. Each of the
representations, $R_n$ (of dimension $n$), appears as an $n{\times}n$
block $n$ times in the decomposition.

This information about the representations (and hence conjugacy
classes) of $\Gamma$ is succinctly encapsulated in the extended Dynkin
diagrams for the associated $A$--$D$--$E$ root systems\McKay\ depicted
in Figure~1.

\topinsert{\vskip1.1cm
\hskip3.5cm\epsfxsize=2.0in\epsfbox{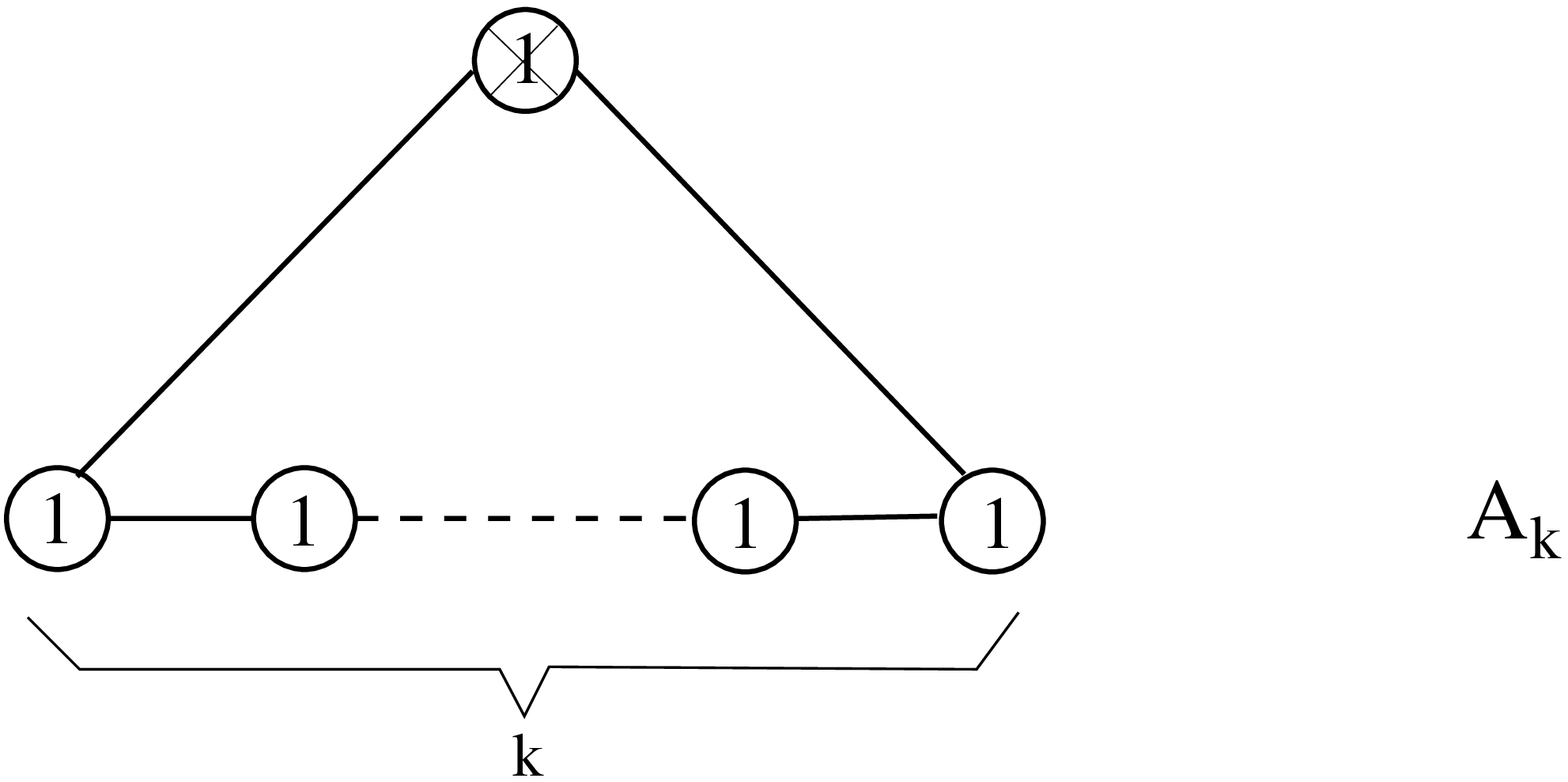}
%\vskip0.1cm

\vskip1.1cm
\hskip3.5cm\epsfxsize=2.1in\epsfbox{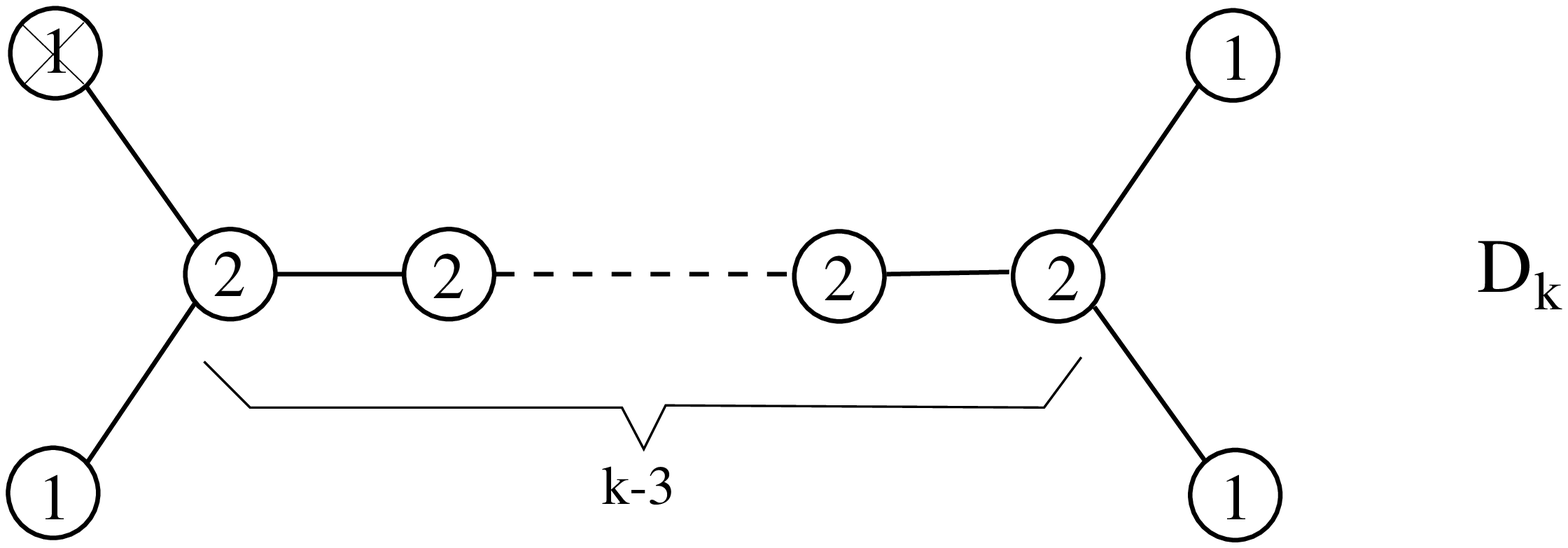}
\vskip1.0cm

\vskip1.1cm
\hskip3.5cm\epsfxsize=2.6in\epsfbox{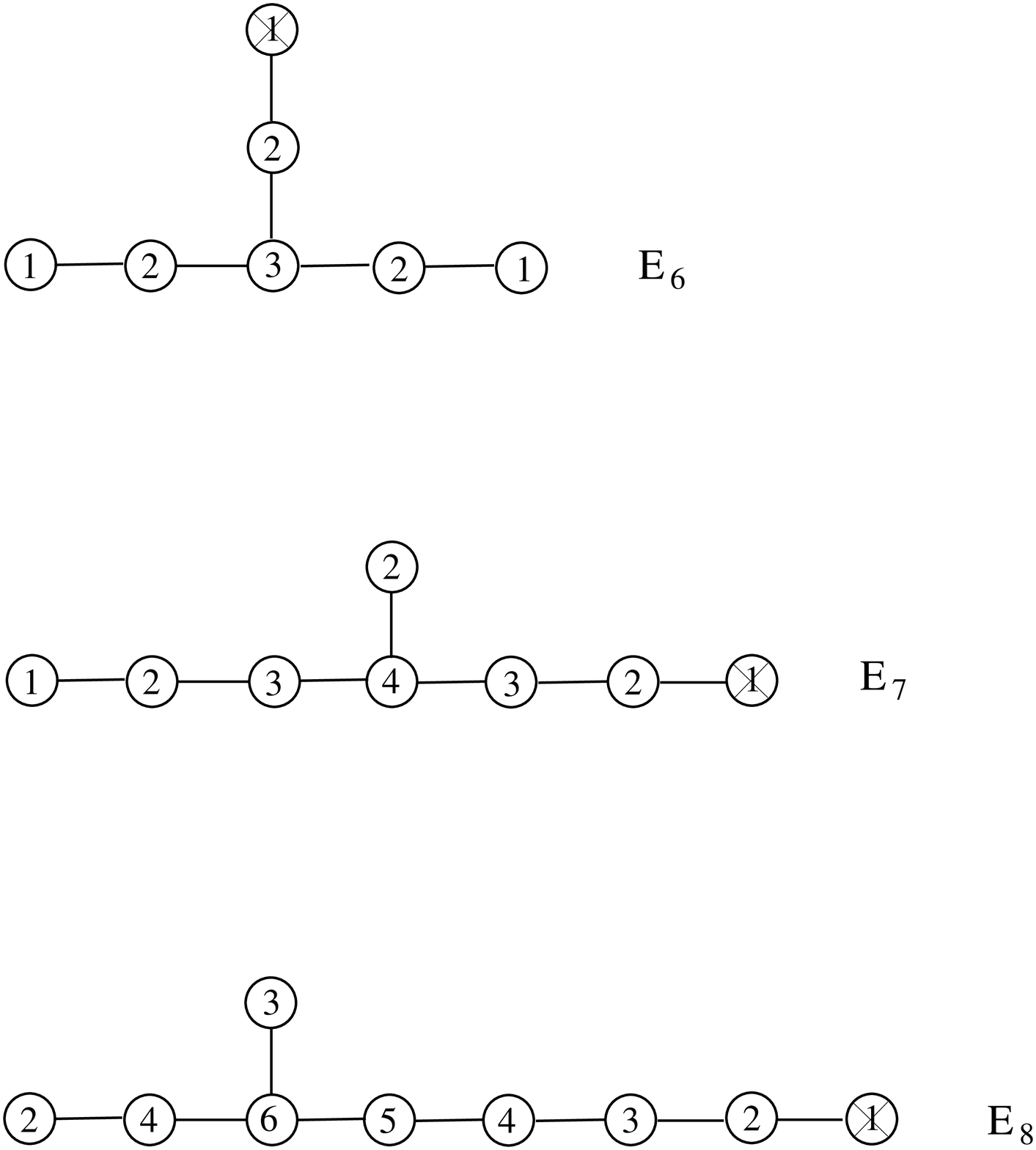}
\vskip0.5cm}
\noindent{{\bf Figure  1.} \sl The Dynkin
 graphs of the extended $A$--$D$--$E$ root systems. They are
isomorphic to the structure of irreducible representations of the
discrete subgroups of $SU(2)$. They encode the arrangement and
resulting spectrum of the open string sectors (D--branes), and also
organise the twisted sectors of the closed string spectrum. See text
for details.}\endinsert

In the Dynkin diagrams, each vertex represents an irreducible
representation of $\Gamma$. The integer in the vertex denotes its
dimension. The special vertex with the `$\times$' sign is the trivial
representation, the one dimensional conjugacy class containing only
the identity. The specific connectivity of each graph encodes the
information about the following decomposition:
\eqn\decompose{Q\otimes R_i=\bigoplus_j a_{ij}R_j,}
where $R_i$ is the $i$th irreducible representation and $Q$ is the
defining two dimensional representation.  Here, the $a_{ij}$ are the
elements of the adjacency matrix~$A$ of the simply laced extended Dynkin
diagrams.

The process of solving equations \vectss\ and \hypers\ to find the
gauge content of our world--volume theory is made much simpler with
the knowledge of these decompositions. In the block--diagonal basis
alluded to above, we can see (for example) that the only nonvanishing
off--diagonal blocks in $\lambda_V$ are those connecting different
copies of a given representation~$R_n$.  In the end, exactly enough
conditions are imposed so as to retain only the content of an
arbitrary $n{\times}n$ Hermitian matrix for each irreducible
representation $R_n$ (repeated $n$ times).

This means that the gauge group is:
\eqn\gauge{F=\prod_i U(n_i)}
where $i$ labels the irreducible representations $R_i$ of dimension
$n_i$.  Pictorially, the gauge group associated with a D--string on a
ALE singularity is simply a product of unitary groups associated to
the extended $A$--$D$--$E$ Dynkin diagram, with a unitary group coming
from each vertex. (See Figure~1.)

Turning to the hypermultiplets, as stated before, we trivially have
dim($F$) hypermultiplets transforming in the adjoint of $F$. These come
from the $2,3,4,5$ sector. Equivalently, they are simply the internal
components of the six dimensional vectors after dimensional reduction.
 
%\filbreak

More interestingly, we have hypermultiplets coming from the $6,7,8,9$
sector. To solve the equation \hypers\ in the block--diagonal basis
for $\gamma^{\phantom{-1}}_\Gamma$ is almost as simple as it was for
the vectors.  The result is best written with a pair of
matrices\foot{As the two dimensional representation $Q$ of $\Gamma$
contains off--diagonal elements in general, components of
$\lambda_H^1$ and $\lambda_H^2$ are related.  The exception is the
case of the $A_k$ series.}, $\lambda_H^1$ and $\lambda_H^2$ in which
non--zero entries appear in off--diagonal blocks which connect
different unitary groups making up $F$. The structure of these
non--zero elements is isomorphic to the adjacency matrix~$A$ of the
extended Dynkin diagram.

In other words, these hypermultiplets transform in the fundamentals of
the unitary groups, according to the representations
\eqn\charges{\bigoplus_i a_{ij}({\bf n}_i,{\bar{\bf n}}_j).}
Pictorially, the hypermultiplets are simply the links of the extended
Dynkin diagrams.

The complete picture with all of the D--strings sitting at the
singular point (the `origin') shows some interesting structure: In
this `blown--down' limit of the ALE space, the $\Gamma$~projection
arranges the $|\Gamma|$ D1--brane probes on the vertices of the
associated extended Dynkin diagram for $A$, $D$ or $E$ simple groups,
depending upon $\Gamma$. On the $i$th vertex labelled $n_i$, there are
$n_i^2$ D1--branes. There are massless vector-- and hyper-- multiplets
in the D1--brane world--volume theory arising from the massless
fundamental strings which connect the various branes.

The world--volume  spectrum is summarised in  Table~1.
\topinsert{
\bigskip
\vbox{
$$\vbox{\offinterlineskip
\hrule height 1.1pt
\halign{&\vrule width 1.1pt#
&\strut\quad#\hfil\quad&
\vrule#
&\strut\quad#\hfil\quad&
\vrule#
&\strut\quad#\hfil\quad&
\vrule width 1.1pt#\cr
height3pt
&\omit&
&\omit&
&\omit&
\cr
&\hfil $\Gamma$&
&\hfil Vector--multiplets&
&\hfil Hyper--multiplets~II&
\cr
height3pt
&\omit&
&\omit&
&\omit&
\cr
\noalign{\hrule height 1.1pt}
height3pt
&\omit&
&\omit&
&\omit&
\cr
&\hfil $\IZ_{k+1}$ &
&\hfil $U(1)^{k+1}$&
&\hfil $\{k{+}1\}{\times}{\bf(1,1)}$ &
\cr
height3pt
&\omit&
&\omit&
&\omit&
\cr
\noalign{\hrule}
height3pt
&\omit&
&\omit&
&\omit&
\cr
&\hfil $\ID_{k-2}$&
&\hfil $U(1)^4{\times}U(2)^{k-3}$&
&\hfil $4{\times}{\bf(1,2)}{+}\{k{-}4\}{\times}{\bf(2,2)}$&
\cr
height3pt
&\omit&
&\omit&
&\omit&
\cr
\noalign{\hrule}
height3pt
&\omit&
&\omit&
&\omit&
\cr
&\hfil $\cal T$&
&\hfil $U(1)^3{\times}U(2)^3{\times}U(3)$&
&\hfil $3{\times}{\bf(1,2)}{+}3{\times}{\bf(2,3)}$&
\cr
height3pt
&\omit&
&\omit&
&\omit&
\cr
\noalign{\hrule}
height3pt
&\omit&
&\omit&
&\omit&
\cr
&\hfil $\cal O$&
&\hfil $U(1)^2{\times}U(2)^3{\times}U(3)^2{\times}U(4)$&
&\hfil $2{\times}\{{\bf(1,2)}{+}{\bf(2,3)}{+}{\bf(3,4)}\}{+}{\bf(2,4)}$&
\cr
height3pt
&\omit&
&\omit&
&\omit&
\cr
\noalign{\hrule}
height3pt
&\omit&
&\omit&
&\omit&
\cr
&\hfil $\cal I$&
&\hfil 
$U(1){\times}U(2)^2{\times}U(3)^2{\times}U(4)^2{\times}U(5){\times}U(6)$&
&\hfil $\eqalign{&{\bf (1,2) }{+}{\bf(2,3) }{+}{\bf(3,4) }{+}{\bf(4,5) }{+}\cr
&{\bf (5,6)}{+}{\bf(2,4) }{+}{\bf(4,6) }{+}{\bf(3,6)}}$&
\cr
}
\hrule height 1.1pt}
$$
}
\bigskip
\noindent{{{\bf Table 1:} \sl
The open string spectrum of the D--brane world--volume theory. It is
an $\N{=}4$ supersymmetric gauge theory in $D{=}2$. In addition, the
hyper--multiplets I transform in the adjoint of the gauge
group $F$ appearing in the vector--multiplets column. 
See text for details.}}}
\endinsert

\def\tphi{\tilde{\phi}}

\subsec{\sl The Moduli Space of Vacua: Part I}
So on the two dimensional world--volume of the D--strings,
we have an $\N{=}4$ gauge
theory with gauge group~$F$ given in the  table.  The
hypermultiplets transform under the gauge symmetry as
\eqn\transform{\delta\phi_H=i\varepsilon_a[\lambda^a_V,\phi_H] \qquad\quad
\delta\psi_H=i\varepsilon_a[\lambda^a_V,\psi_H]} 
and the explicit form of the $\lambda$--matrices yields the charges
under $F$ described in the table. The diagonal $U(1)$ gauge group acts
trivially in all cases, and so the non--trivial gauge group actually
is $F/U(1)$.  Similarly there is a hypermultiplet, which we will
distinguish as $\tphi_H$, with a Chan--Paton matrix proportional to
the identity and hence neutral with respect to the entire gauge group.
 
It is worth noting here that this physical structure has already been
presented in the mathematics literature\kronheimer\ in an existence
proof for the metrics on the ALE spaces.  In ref.\kronheimer, there is
a space $M$, a flat hyperK\"ahler manifold, defined as the
$\Gamma$--invariant subspace of a space $P{=}Q{\otimes}{\rm
End}(R)$. $R$ is the regular representation of $\Gamma$ and $Q$ is the
defining two--dimensional representation.  There is a group of unitary
transformations $U(|\Gamma|)$ naturally acting on $R$, and hence also
on the space $P$ by construction.  The subgroup of these
transformations which commutes with the action of $\Gamma$ on $P$ is a
group denoted $F$. The group $F/U(1)$ acts non--trivially on the
manifold $M$, preserving the three complex structures $\cal I,\cal J$
and $\cal K$ of the hyperK\"ahler manifold.

Returning to the physics, we see that remarkably these mathematical
tools have a clear physical origin!  The space $P$ is exactly
equivalent to our space of $\psi_H$--hypermultiplets before
projection, and the restriction to $M$ is the projection \hypers. The
quaternionic space $M$ is the space of hypermultiplets displayed in
the table. These hypermultiplets, with their content of four scalars
each, naturally display the quaternionic structure crucial to the
construction of ref\kronheimer.  Physically the origin of this
quaternionic structure lies in the form appearing in the
four-dimensional background $\IR^4$ (see equation \quater). The
positions of the D1--branes in this space are described by these
hypermultiplets.  Notice that the number of hypermultiplets is equal
to $|\Gamma|$.  There are also $|\Gamma|$ D1--branes present at the
origin of the singular space. The gauge group $F$ corresponds
precisely to the unitary transformations appearing in the construction
of ref\kronheimer.  (Indeed, we use the same notation, for
simplicity.)

The scalar potential of our $D{=}2$, $\N{=}4$ world--volume theory can
be written as
\eqn\nooterms{\sum_{a,b}{\rm Tr}|[\psi_H^a,\psi_H^b]|^2
+2\sum_{a,i}{\rm Tr}|[\psi_H^a,\phi_H^i]|^2+
\sum_{i,j}{\rm Tr}[\phi_H^i,\phi_H^j]^2}
where in these sums $a,b=1,2,\bar{1},\bar{2}$ 
and $i,j=2,3,4,5$. Here, this commutator structure is inherited
through the dimensional reduction (and $\Gamma$--projection) of the
ten--dimensional $U(|\Gamma|)$ Yang-Mills action.

Broadly speaking, there are two very distinct branches of the moduli
space of vacua of the theory\foot{We are grateful to M. Strassler for
a conversation which helped to clarify the field theory
terminology.}. That with $\psi_H{=}0$ and $\phi_H{\neq}0$
is the `Coulomb' branch,
and the branch with $\psi_H{\neq}0$ and $\phi_H{=}0$, the `Higgs' branch
\foot{It is interesting that in the
orbifold limit, there are no intermediate situations with both
$\psi_H{\neq}0$ and $\phi_H\neq0$.  This will no longer hold when the
singularity is resolved as in the following section.}.

The Coulomb branch is the branch of moduli space where the gauge
symmetry is generically $U(1)^r$, where $r{=}{\sum_i n_i}$ and $n_i$
is the dimension of the $i$th irreducible representation, $R_i$,
of~$\Gamma$. This represents a family of D1--branes all living on the
singular orbifold point in $\IR^4/\Gamma$ while moving independently
in the $x^2,x^3,x^4,x^5$ directions\foot{Actually a complete
exploration of the Coulomb branch would lead us to singularities, the
discussion of which will take us beyond the scope of this paper. A
complete discussion involves the study of the scalars coming from the
closed string sector (see later). These include a family of
theta--angles $\Theta_i$, one for every $U(1)$ in the gauge group, and
families of other scalars. In leaving them out of the discussion so
far, we have tacitly assumed that we have set them all to zero. In
fact, it has been shown that the orbifold retains non--zero values for
the theta--angles\aspinwalltwo. It is sufficient to avoid the
singularities in the Coulomb branch this way by keeping\edcomm\ the
$\Theta_i{\neq}0$. Tuning the theory further by setting all of the
closed string scalars to zero takes us to regions of moduli space
containing singularities which are of interest ({\it e.g.}, the
strings become tensionless), but are not the subject of this paper.}.
This generic situation comes from giving non--zero expectation values
to the scalars $\phi_H$ in the Cartan subalgebra of Lie($F$) (ensuring
that the third term in the potential \nooterms\ vanishes). This
produces mass terms for the vectors filling out the gauge group $F$,
and breaking it to $U(1)^{r}$. Similarly most of the scalars
become massive leaving massless those $\phi_H$ carrying only Abelian
charges, as well as the neutral $\tphi_H$.  These scalars correspond
to the D1--brane positions in the $x^2,x^3,x^4,x^5$
directions\foot{Note that with the $A_k$ series, for which the gauge
group $F$ is entirely Abelian, on the Coulomb branch we are giving
expectation values to all of the hypermultiplets $\phi_H$ and
$\tphi_H$, and none of the gauge symmetries are broken.}.

The Higgs branch will be the main interest for us in the rest of the
paper as it is the branch concerned with the resolution of the
singularity. In the generic situation when we give expectation values
to the hypermultiplets in $\psi^{1,2}_H$, we will Higgs away most of
the gauge group. Naively, as the dimension of the gauge group is also
$|\Gamma|$, it may seem that we can Higgs everything away. This is not
quite correct, as it is the group $F/U(1)$ which acts non--trivially
on the hypermultiplets. So the diagonal $U(1)$ remains unbroken, and
as well the corresponding $\tphi_H$ multiplet remains massless. The
unbroken $U(1)$ is the familiar gauge group for a single brane, and
the $\tphi_H$ hypermultiplet encodes its position in the
$x^2,x^3,x^4,x^5$ directions. Similarly the space of allowed values
for the $\psi^{1,2}_H$ hypermultiplets corresponds to the positions
that the single D1--brane can occupy in the $x^6,x^7,x^8,x^9$
directions.

To see that the allowed space in which the D1--brane can propagate is
indeed four dimensional is relatively easy.  On this branch of moduli
space, asking that the potential
\nooterms\ vanishes for generic $\psi^{1,2}_H{\neq}0$ will first of
all require $\phi_H{=}0$ except for the uncharged hypermultiplet ({\it
i.e.,} that with a Chan-Paton matrix proportional to the
identity). With some algebra, one can rewrite the first term in the
potential as a sum of three D--terms
\eqn\nodterms{\left({\rm Tr}
\left[\lambda^a_V\cdot\left\{\Psi^{1\dagger}_H {\bf\Sigma}\Psi^1_H
+\Psi^{2\dagger}_H {\bf\Sigma}\Psi^2_H\right\}
\right]\right)^2}
where 
\eqn\quatern{\Psi^{1\dagger}_H=\left(\psi^{1}_H,-\psi^{2\dagger}_H\right)
\quad{\rm and}\quad
\Psi^{2\dagger}_H=\left(\psi^{2}_H,\psi^{1\dagger}_H\right) } 
are the natural $SU(2)_R$ doublets appearing in the quaternionic form
\hyperquat. The $\lambda^a_V$ are generators of $F/U(1)$. The three 
components, $\sigma^i$, of ${\bf\Sigma}$
are the Pauli matrices acting on the $SU(2)_R$ doublet space.  Thus
the vanishing of the individual D--terms is a set of $3(|\Gamma|{-}1)$
real constraints on the available $4|\Gamma|$ dimensional space
parameterised by the hypermultiplets.  This restriction gives us a
space of overcounted vacua, since we have not accounted for the gauge
symmetries.  So we must impose another $|\Gamma|{-}1$ gauge-fixing
conditions, which leaves us with a space of four real dimensions.

For these vacua on the Higgs branch then, the D1--branes have moved
off the orbifold point, leaving a single D1--brane in open space.
Equivalently, we can say that the $|\Gamma|$ D1--branes are now all
away from the origin of $\IR^4$, but that the $\Gamma$--projection
relates them as images of one another, leaving only one independent
D1--brane.

There is a natural way to characterise this space of vacua
algebraically\algebra. It is possible to determine a family of
polynomials of the coordinates $z^1$ and $z^2$ which are invariant
under the action of the group $\Gamma$ on the space $\IC^2$. In each
case ($A$, $D$ or $E$) there are three such distinguished invariants,
$x, y$ and $z$.  These invariants satisfy a well known polynomial
relation in each case, ${\cal W}(x,y,z){=}0$, where
\eqn\poly{\eqalign{{\cal W}_{A_k}&=xy+z^{k+1};\cr
{\cal W}_{D_{k}}&=x^2+y^2z+z^{k-1};\cr
{\cal W}_{E_6}&=x^2+y^3+z^4;\cr{\cal W}_{E_7}&=x^2+y^3+yz^3;\cr
{\cal W}_{E_8}&=x^2+y^3+z^5.}}
Algebraically, the singular ALE space $\IC^2/\Gamma$ is described
as a variety ${\cal V}_0$ in $\IC^3$ described by the vanishing locus
${\cal W}(x,y,z){=}0$, where $x,y$ and $z$ are coordinates on
$\IC^3$. 

 The space of hypermultiplets $\psi^{1,2}_H$, as acted on by the gauge
group $F$, inherits much of the structure just described. Indeed, the
procedure of finding the gauge invariant family of allowed vacua
defines three combinations of hypermultiplets, $X,Y$ and $Z$ which are
isomorphic to the $x,y$ and $z$ of the algebraic
discussion\foot{This is a result of
geometric invariant theory\mumford.}.

In this way we make our first contact here with a large body of
classic results concerning the mathematical nature of the
$A$--$D$--$E$ singularities. Also, we have actually described a
special case of the `HyperK\"ahler Quotient' technique\rocek, to
recover the singular $\IC^2/\Gamma$ spaces. The next step is to
introduce the remaining elements of the construction ---arising in the
closed string sector--- which will allow us to construct the deformed
spaces, complete the description of the hyperK\"ahler quotient, and
discuss metrics on the ALE spaces.

\newsec{D--Strings on ALE Spaces: The Blow--Up}

The closed string sector provides the remaining fields of relevance.
Our discussion in the previous section was restricted to the case
when their expectation values were set to zero.

\subsec{\sl The Closed String Sector}
The closed string spectrum is much more familiar in this context. The
traditional route to the spectrum is via orbifold techniques.

In general, the orbifold procedure is complicated somewhat by the
non--Abelian nature of the group $\Gamma$ (only for the $A_k$ series is
it Abelian). Using the naive twisted sectors would not lead to a
modular invariant theory, as sectors twisted by elements in the same
conjugacy class can mix\orbifold. The procedure for computing twisted
sectors thus takes into account the structure of the conjugacy classes
discussed above. Each conjugacy class ultimately contributes a
spacetime field with well--defined transformation properties under the
Lorentz group.

We are interested in that part of the closed string spectrum which
consists of the scalars corresponding to the moduli we use to blow up
the orbifold. We can identify these scalars quite quickly using
knowledge of the algebraic resolution of the ALE spaces.

In the resolved space, there is an algebraic description (see, for
example, ref.\algebra) of the region which will become the singular
point in the blow--down in terms of a family of two--cycles. We can
deduce what scalars arise in the closed string theory by contracting
the various rank two tensor fields on these cycles. These will be the
fields which arise in the orbifold limit if we chose to enumerate the
spectrum directly.

Deformation of ${\cal V}_0$ is done via elements of the ring of
polynomials ${\cal R}{=}\IC^3[x,y,z]/\partial{\cal W}$. Just away from
the singular limit, this deformation defines a map $\rho$ from the
smooth `minimally resolved' variety $\cal V$ to the singular variety
${\cal V}_0$ which is an isomorphism everywhere except at the singular
point itself, $\{0\}$.  The neighbourhood which maps to the singular
point, $\rho^{-1}({\cal V}_0{-}\{0\})$, (the `exceptional divisor')
is described as a series of algebraic curves $\IP^1$, or two--cycles,
${\bf c}_i$ (essentially two--spheres $S^2$) which have an
intersection matrix which is the (negative) Cartan matrix: ${\bf
c}_i\cdot {\bf c}_j{=}-2\delta_{ij}{+}a_{ij}$, where $a_{ij}$ is the
Dynkin adjacency matrix we encountered earlier. Here, $i$ and $j$ run
over only the ($n{-}1{=}k,k,6,7$ or $8$) non--trivial representations
of $\Gamma$, {\it i.e.}, there is no two--cycle ${\bf c}_0$
corresponding to the extended vertex of the Dynkin graph.  Therefore,
the two--cycles have the structure of the Dynkin diagrams we found
earlier:  Each non--trivial conjugacy class has a two--cycle
associated with it.

The ten dimensional type~IIB theory contains the following rank two
tensors: There is the metric $G$ and the antisymmetric tensor field
$B^{(2)}$, from the NS-NS sector.  The R-R sector supplies the two--form
$A^{(2)}$.  Therefore, from each conjugacy class, we get (in the flat six
dimensions) two scalars (corresponding to $\theta$--angles) arising
from contracting the forms $A^{(2)}$ and $B^{(2)}$ with the two--cycles.
Meanwhile, three scalars come from the metric $G$. This is the scalar
content of a tensor multiplet of ${\cal N}{=}2$ supersymmetry\foot{We
denote the transformation properties under the $SU(2){\times}SU(2)$
little group.} in $D{=}6$: ${\bf(1,3)}{+}5{\bf(1,1)}$.
The self--dual antisymmetric tensors in these multiplets arise from
contracting the R-R sector self--dual four--form $A^{(4)}$ on the
two--cycles.
In this way, each non--trivial conjugacy class gives rise to a
complete tensor multiplet.

Another, perhaps more contemporary way of seeing how to obtain this
result is to realise that those twist fields are precisely the set of
closed string fields which couple to the various string solitons
arising from wrapping the self--dual D3--brane of ten dimensional IIB
theory on the two--cycles. (See ref.\edcomm\ for the $A_1$ case of
this.)  Here, this reduction procedure gives rise to $n{-}1$ different
types of D--string in the theory, 
plus one more type corresponding to the `pure' D1--brane ({\it
i.e.,} the familiar string soliton which couples directly to $A^{(2)}$).  

Continuing the organisation of our physics by the
representation theory of $\Gamma$, we see that there is exactly one
species of D--string in the problem for every conjugacy class.
Returning to the discussion in section 2.4, it is natural that the
$|\Gamma|$ D1--branes living on the fixed point in the Coulomb branch
can be labelled according to which species they belong to, resulting
in multiplicities which furnish a physical version of the `regular'
$|\Gamma|$--dimensional representation, $R$, of $\Gamma$.
 
This arrangement is the minimum requirement for there to be a flat
direction in the potential corresponding to moving off the fixed point
and into open space $\IR/\Gamma$. The~$|\Gamma|$ strings coalesce into
one string which carries zero charge under the twisted sector fields.
This single string is to be identified with the `pure' D1--brane, as
the other types of string cannot exist in isolation off the fixed
point given that that they couple to  twisted sector closed
string fields, which are localised there. As mentioned earlier it
would be consistent to place more D--strings on the singularity, but
only multiples of $|\Gamma|$ can move off the
singularity\refs{\ericjoe,\ericmeI}, in the manner just described.

Of the twisted sector tensor multiplet fields which we just discussed,
it is the trio of NS-NS scalars which will interest us, for the
purposes of performing the blow--up.  They transform as a triplet
under the~$SU(2)_R$ symmetry acting on the $x^6,x^7,x^8,x^9$ space.
We shall denote them ${\bf D}^i$, the label~$i$ running over the
different conjugacy classes ({\it i.e.,} the vertices in the
(unextended) Dynkin diagram).  Here the $SU(2)_R$ symmetry has become
the R--symmetry of the $D{=}2$, $\N{=}4$ world--volume theory (see
section 2.1).

\subsec{\sl D--Terms, D--Flatness and the Moment Map}
The closed string fields ${\bf D}^i$ couple into the world--volume
theory via Fayet--Illiopoulos (FI) terms\moore.  The R--symmetry
$SU(2)_R$ of the $D{=}2$ theory with $\N{=}4$ supersymmetry requires
that the D--terms appear in triplets as in equation \nodterms.  Gauge
invariant FI--terms could be written for every $U(1)$ in the
theory. However from the closed strings there is a trio of NS-NS
scalars ${\bf D}^i$ for every non--trivial conjugacy class of
$\Gamma$, and each class is associated with a vertex in the Dynkin
diagram (the extended vertex is not included here). Hence an FI--term
appears only for the $U(1)$ subgroups of each unitary group appearing
for every vertex of the Dynkin diagram ({\it i.e.}, each of the
$U(1)$'s in $F/U(1)$).

The relevant terms in the potential \nodterms\ containing the FI--terms
are 
\eqn\dterms{\left({\rm Tr}
\left[\lambda^i_V\cdot\left\{\Psi^{1\dagger}_H {\bf\Sigma}\Psi^1_H
+\Psi^{2\dagger}_H {\bf\Sigma}\Psi^2_H\right\} \right]-{\bf
D}^i\right)^2,} where the $\lambda^i_V$ are generators of the $U(1)^r$
subgroup of $F/U(1)$ where $r{=}k,k,6,7$ or 8, depending upon
$\Gamma$.  Of course, the terms in \nodterms\ corresponding to the
generators $\lambda^a_V$ not in the Abelian part of the group $F/U(1)$
are still present unchanged in the full potential.

Again, this structure appears in the mathematics literature.  The
existence of a triholomorphic\foot{Here, `triholomorphic' simply
means that it preserves the three complex structures $\cal I,\cal J$
and~$\cal K$ associated with the hyperK\"ahler structure of $M$.}\
symmetry ($F/U(1)$) acting on the hyperK\"ahler manifold $M$
guarantees the existence of a map ${\mu}\hskip-.55em{\mu}$ from $M$ to
$\IR^3{\otimes}{\rm Lie}(F/U(1))$.  This map is called the `moment
map'. The $\IR^3$--valuedness is simply the projection onto the three
complex structures $\cal I,\cal J$ and~$\cal K$ of the manifold $M$,
and so ${\mu}\hskip-.55em{\mu}$ has three components which can be
arranged as a vector of~$SU(2)_R$. The moment map can be written
as\kronheimer:
\eqn\moment{\eqalign{&\mu_1(\psi^1_H,\psi^2_H)=
\phantom{i}[\psi^2_H,\psi^1_H]
+\phantom{i}[\psi^{1\dagger}_H,\psi^{2\dagger}_H]\cr
&\mu_2(\psi^1_H,\psi^2_H)=i[\psi^2_H,\psi^1_H]
-i[\psi^{1\dagger}_H,\psi^{2\dagger}_H]\cr
&\mu_3(\psi^1_H,\psi^2_H)
=\phantom{i}[\psi^1_H,\psi^{1\dagger}_H]+
\phantom{i}[\psi^2_H,\psi^{2\dagger}_H]
}} Hence we have ${{\mu}\hskip-.55em{\mu}{=}\Psi^{1\dagger}_H
{\bf\Sigma}\Psi^1_H +\Psi^{2\dagger}_H {\bf\Sigma}\Psi^2_H}$.
As defined, ${\mu}\hskip-.55em{\mu}$ is a vector with components
in Lie($F/U(1)$) 
which we can project onto a chosen basis.
In \nodterms\ and \dterms, with the trace 
we project it explicitly onto the basis vectors $\lambda^a_V$.

\subsec{\sl The Moduli Space of Vacua: Part II}
As stated earlier, the Higgs branch of vacua concerns us 
here.\foot{Note that for generic values of the ${\bf D}^i$
solving the following two equations will lead to setting
$\phi_H{=}0$ in order that the full potential vanish. However
for a partial resolution of the singularity in which some of
the ${\bf D}^i$ vanish, the vacua may have both $\psi_H\neq0$
and $\phi_H\neq0$.} 
These vacua are now characterised by the equations:
\eqn\nodtermstwo{{\rm Tr}
\left[\lambda^a_V\cdot\left\{\Psi^{1\dagger}_H {\bf\Sigma}\Psi^1_H
+\Psi^{2\dagger}_H {\bf\Sigma}\Psi^2_H\right\} \right]=0} 
and
\eqn\dtermstwo{{\rm Tr}
\left[\lambda^i_V\cdot\left\{\Psi^{1\dagger}_H {\bf\Sigma}\Psi^1_H
+\Psi^{2\dagger}_H {\bf\Sigma}\Psi^2_H\right\} \right]={\bf
D}^i.}  
This set of equations has a natural interpretation in the mathematics
literature.  The hyperK\"ahler quotient construction\rocek\ involves
a choice of numbers $\zeta\hskip-.45em\zeta^i{\in}\IR^3{\otimes} \cal
Z$, where $\cal Z$ is the center of the Lie algebra in which the
moment map $\mmu$ takes is values (Lie($F/U(1)$) for us). The space
$\mmu^{-1}(\zeta\hskip-.45em\zeta^i)\subset M$ is an invariant
submanifold under $F$, by virtue of the fact that the center $\cal Z$
is defined as the set of F--invariant elements. The quotient space
${\cal M}=\mmu^{-1}(\zeta\hskip-.45em\zeta^i)/(F/U(1))$ is the
hyperK\"ahler quotient obtained from $F$ and $M$. In ref.\hitchin, it
is proven that $\cal M$, as obtained in this way, is indeed
hyperK\"ahler.

Our equations \nodtermstwo\ and \dtermstwo, characterising the Higgs
branch of the moduli space are simply equations telling us the
restriction of the hypermultiplet space $M$ to the subspace
$\mmu^{-1}(\zeta\hskip-.45em\zeta^i)$. The $\zeta\hskip-.45em\zeta^i$
parameterising the center of Lie($F/U(1)$) are simply the closed
string scalar fields ${\bf D}^i$ which enter into the construction via
FI--terms.

The rest of the construction continues as follows. By imposing
the potential vanishing conditions we have restricted ourselves to an
$F/U(1)$ invariant subspace of $M$, our flat space of hypermultiplet
values. 
Finally we restrict ourselves to the vacua
which are not related by gauge transformations. This
completes for us the hyperK\"ahler quotient.

To make sure that the counting works out, we can check that the right
number of conditions have been imposed:
The real dimension of the space $M$ is 4$|\Gamma|$
which is four times the number of hypermultiplets.
The moment map is valued in
$\IR^3{\otimes}{\rm Lie}(F/U(1))$ and  the dimension of $F$ is $|\Gamma|$,
so imposing the D--flatness conditions gives 3${\times}(|\Gamma|{-}1)$
conditions. The final gauge--fixing 
restricts us with another $|\Gamma|{-}1$
conditions, leaving a $4|\Gamma|{-}4(|\Gamma|{-}1){=}4$ dimensional space.

This four dimensional hyperK\"ahler space is the resolved ALE
manifold. We see explicitly how the amount of the resolution is
controlled by the expectation values of the closed string twisted
sectors fields, ${\bf D}^i$, a fact that we knew algebraically from
orbifold techniques.

To summarise, we now have a clear understanding of
the mechanics of how fundamental type~IIB string theory probes the
metric geometry of an orbifold, as the problem maps (under strong/weak
coupling duality) to one of finding vacua of a field theory associated
to a D1--brane's world sheet. The metric on the field theory's space
of vacua, parameterised by the hypermultiplets, is precisely the
spacetime metric that the D1--brane sees as it moves around in the ALE
space.

\bigskip
\bigskip

\newsec{On Finding Explicit Metrics}
In the previous sections we have seen the mechanism by which we can
directly extract the details of the geometry of the space that strings
propagate in.

The 4$|\Gamma|$--dimensional metric on the 
(unconstrained) space of  hypermultiplets is simply 
\eqn\metric{ds^2_M=
\Tr\left\{d\psi^{1\dagger}_Hd\psi^1_H
+d\psi^{2\dagger}_Hd\psi^2_H\right\}.} The next natural step is
to find the metric on the gauge invariant submanifold. This is done by
replacing the derivatives `$d$' above by covariant derivatives,
minimally coupling the $\psi_H$ to the two dimensional gauge fields
$A{=}A_a\lambda^a_V$:
\eqn\minimal{d\psi_H\to D\psi_H= d\psi_H+iA_a[\lambda^a_V,\psi_H].}
The coset metric can now be obtained by simply choosing a gauge and
integrating out\foot{In general, there are of course $\alpha^\prime$
(inverse D--string tension) corrections to this saddle point
procedure. However there are no corrections from higher orders in
string coupling.  There are two ways to see this {\it (i)} In the
equivalent $D{=}6$, $\N{=}1$ system, the dilaton arises in a tensor
multiplet, which becomes a vector multiplet in lower dimensions. As
our metric is on the hypermultiplet moduli space, string loops
(controlled by the dilaton) do not affect it; {\it (ii)} The
strong/weak self--duality of the theory assures us that since the
background metric which the fundamental string sees is not
corrected at higher order in $\alpha^\prime$ (it's a $D{=}2$, $\N{=}4$
nonlinear sigma model), then the metric seen by the D--string is not
corrected by string loops.}  the gauge fields $A$.
After imposing the $3(|\Gamma|{-}1)$ constraints
from the moment map, we have the final metric in which the
($r{=}k,k,6,7$ or $8$) arbitrary
vectors ${\bf D}^i$ appear as parameters. The freedom to make different
gauge choices translates into (a subset of) the freedom to make
coordinate redefinitions in the final metric. Failure to completely
fix the gauge will result in $N$ redundant coordinates in the metric
where $N$ is the dimension of the {\it continuous} subgroup left
unfixed.  We stress continuous because there is always the possibility
that there is some {\it discrete} subgroup of the gauge symmetry left
unfixed. Then there will be no redundant coordinates, but it will be
necessary to make discrete identifications on the space defined by
the final choice of coordinates. Notice that the most natural discrete
subgroup of $F/U(1)$ is $\Gamma$. We expect that the discrete
identification by~$\Gamma$ present in the ALE spaces will arise in the
explicit metrics by precisely such an incomplete gauge fixing. (See
ref.\myoldpaper\ for such a coset example.)

In this way, we can in principle write down all of the metrics on the
ALE spaces. Indeed, this is by now a well--known exercise for the
Eguchi--Hanson metric, and it was explicitly carried out in this
context in ref.\joetensor\ for the $A_k$ series.  In this case, a
number of fortuitous occurrences make the problem relatively
straightforward: 

For example, the Abelian nature of the problem translates into a
simple structure for the relations imposed by the moment map: A basis
can be found in which there is a {\it gauge invariant} combination of
hypermultiplets of the form
$(\psi^{\dagger}_H{\bf\Sigma}\psi^{\phantom{\dagger}}_H)^i$,
associated to the $i$th vertex in the $A_k$ Dynkin diagram, as the
superscript denotes. The $k{+}1$ moment map relations are then simply
of the form:
$(\psi^{\dagger}_H{\bf\Sigma}\psi^{\phantom{\dagger}}_H)^i-(\psi^{\dagger}_H
{\bf\Sigma}\psi^{\phantom{\dagger}}_H)^{i-1}={\bf D}^i$, where
$i{=}1,\ldots,k{+}1$, reflecting the cyclic structure of the extended
Dynkin graph for $A_k$.

Further to this, the `integrating out of the gauge fields' procedure
produces a metric which is already partially written in terms of those
same gauge invariant combinations of hypermultiplets in terms of which
the moment map is written. This makes straightforward the step of
fixing a gauge ({\it i.e.,} there is no need to) and finding
coordinates simultaneously adapted to the metric and the subsequent
substitution of the moment map constraints into
it. Specifically\joetensor, to obtain the standard form \Ametric, the
coordinate ${\bf
y}{=}(\psi^{\dagger}_H{\bf\Sigma}\psi^{\phantom{\dagger}}_H)^0$, and
the coordinates of the multicenters are ${\bf y}_i=\sum_{a=1}^i {\bf
D}^a$.

Even without those pleasant structures, the problem is
made easier by the fact that the simplest example $A_1$ is relatively
easy to handle algebraically (the starting metric is only 8
dimensional), and so it would be possible to work by hand to discover
the gauge and coordinate choices had they been difficult to find. The
cyclic nature of the Dynkin diagrams means that those choices then
generalise easily to the full $A_k$ series, as above.

Turning to the $D$ and $E$ series, it is easy to see why there are
currently no (to our knowledge) closed expressions for their metrics
in the literature!  Hardly any of the simplifying facts mentioned
above appear to be true in these cases, as one finds after some
exploration.  The moment map does not suggest an obvious set of
choices for gauge invariant coordinates on the final
space. Furthermore, it is not clear what gauge slice to choose, in the
absence of such coordinates. From experience with the $A_k$ series,
one might expect that there might be clues from the metric one gets
after integrating out the gauge fields. This may be true, but given
that for the simplest example, $D_4$, that metric is 25
dimensional\foot{To obtain this metric requires an inversion of a
$7{\times}7$ matrix of algebraic expressions. This already a large
amount of algebra.}, it is a daunting task to find the correct choices
by eye. However, it might turn out that there is some symmetry (such
as $SU(2)_R$) which might serve as guidance, organising the algebra
and help solve the problem. One might hope that the coordinates would
then generalise to the full $D_k$ series, once found\foot{In this
instance, it might be that $D_4$ is too simple a first example, as it
has none of the $(2,2)$ hypermultiplets, the main structure which
generalises in the higher $k$ examples. Note however that starting
metric of $D_5$ is 48 dimensional!}.

The astute reader by now should have begun to suspect one of two
possibilities: Either {\sl(i)} we have managed to surpass all of the
difficulties described above, and are only emphasizing the magnitude
of the problem to make the presentation of an elegant solution more
dramatic, or {\sl(ii)} we have failed to find the explicit form of the
metrics, and are reporting our observations in the hope that they may
help others interested in the problem.

Unfortunately, it is the latter {\sl(ii)} which is the case. Although
perhaps only of aesthetic interest, finding closed forms for the
metrics generalising \Ametric\ is an intriguing problem, and we still
hope that a solution can be found in the near future.

\newsec{Closing Remarks}
This addition of D--brane technology and strong/weak coupling duality
to traditional closed string methods in order to understand
non--trivial string theory backgrounds is quite satisfying.

The duality map allowed us to relate the question of how strings probe
the geometry of the ALE spaces to the study of vacua of two
dimensional ${\cal N}{=}4$ supersymmetric field theories on D1--brane
world--volumes. The appropriate two dimensional theories derived here
had a spectrum directly related to the work of Kronheimer.
 
The discussion presented in this paper was restricted to the vacua of
two dimensional field theories because the probes were D1--branes.  In
ref.\ken, a study is presented of the vacua of ${\cal N}{=}4$
supersymmetric {\it three dimensional} field theories with the
(Kronheimer) spectrum discussed here\foot{We are grateful to
K. Intriligator and M. Strassler for bringing this paper to our
attention.}. Although the context of that paper is field theory, one
should be able to obtain a D--brane realisation of that scenario {\it
via} T--dualising our present situation along the $x^2,x^3,x^4$ or
$x^5$ coordinate. This will result in type~IIA theory with D2--brane
probes, whose three dimensional world--volumes will provide a natural
setting for the discussion of ref.\ken. The subsequent interpretation
of the `mirror symmetry' duality found in ref.\ken\ for those three
dimensional models should be very interesting.
 
Returning to the two--dimensional (world--sheet) setting of this paper,
the most direct approach to discussing a full (perturbative) string
theory solution representing a closed string propagating in a certain
background is via the conformal field theory on the world--sheet.
However, knowledge of the full conformal field theory is not always a
luxury which is available to us. Furthermore, even the knowledge of a
conformal field theory description of a background at some point in
the moduli space is not always enough to discover important global
aspects of the moduli space of backgrounds. To circumvent these
limitations, Witten studied how we can extract useful information
about the string propagation on a non--trivial background (and often a
whole moduli space of backgrounds) by studying two dimensional gauged
linear sigma models\phases. These linear sigma models do not represent
solutions to the closed string equations of motion ({\it i.e.},
conformal field theories), but are connected to such solutions by
renormalisation group flow to an infra--red fixed point.

Many characteristic properties of the spacetimes we wish to study
(interpreted as properties of the two dimensional world--sheet
theories) can be incorporated easily by hand into the linear sigma
model. Under renormalisation group flow, these properties survive to
become the required properties of the conformal field theory
representing the solution. In this indirect way, we can study many
aspects of these non--trivial string theory backgrounds by
manipulations of the linear sigma model.

It is now clear that strong/weak coupling duality and D--branes shed
new light upon the role of these linear sigma models: They tell us the
precise circumstances under which the gauged linear sigma models, as
world--sheet theories of dual strings {\it are} solutions representing
string theory backgrounds.  The new ingredient is the possibility of
adding open string sectors (D--branes). So duality knows how to move
along the renormalisation group trajectories the existence of which we
relied upon previously.

The question arises as to whether these observations have taught us
anything about the `larger picture' concerning the theory underlying
string theory and eleven--dimensional supergravity.  On the one hand,
we can take the conservative point of view that D--brane probes simply
give us a new means of building linear sigma models, which is already
progress.

On the other hand, let us 
suggest the following
possibility: As strong/weak coupling duality seems to relate different
points on the  renormalisation group flow trajectory,
we might interpret this as a new clue as to the nature of duality
itself, or at least a new handle on it. For every conformal field
theory (representing a closed string background, say), there is a
whole universality class of sigma models which flow to it in the
infra--red. Perhaps for every one of these models, there is implied a
duality transformation to a new theory in which the sigma model
spectrum represents a valid configuration. The new dual theory could
be either the same string theory, a new one, or something else. 

Let us briefly engage in a little revisionist history to make the
point: The knowledge\edcomm\ that the linear sigma model containing
Kronheimer's data flows to the conformal field theory representing the
type~IIB string in an ALE background would imply the existence of a
duality transformation to a new theory where the linear sigma model is
a solution. Researchers examining the sigma model in that light would
then discover that it was the theory of a D1--brane of the same
type~IIB theory on the ALE background. A similar story might be told
for the $SO(32)$ theory: The knowledge\adhm\ that the linear sigma
model with the ADHM data flows to the conformal field theory
representing the $SO(32)$ heterotic string in a Yang--Mills instanton
background would suggest to physicists the existence of a dual theory
giving rise to that linear sigma model. This eventually turns out to
be\douglas\ the $SO(32)$ type~I theory with D1--and D5--branes.

It would certainly be interesting to investigate this suggestion
further.

%%%%%%%%%%%%%%%%%%%%%%%%%%%%%%%%%%%%%%%%%%%%%%%%%%%%%%%%%%%%%%%%%%%%%%%%%%
\bigskip
\medskip
%\vskip0.5truecm

\noindent
{\bf Acknowledgments:}

\noindent
We would like to thank Eric Gimon and Joe Polchinski for helpful
comments.  CVJ was supported in part by the National Science
Foundation under Grant No. PHY94--07194. RCM's research was supported
by NSERC of Canada, Fonds FCAR du Qu\'ebec. RCM would like to thank
the Erwin Schr\"odinger International Institute for Mathematical
Physics for hospitality when this work was initiated.  CVJ would also
like to thank the Physics Department at McGill University for
hospitality while some of this research was carried out.

%\vfill\eject
\listrefs 
\bye